\documentclass[journal]{IEEEtran}

\usepackage{cite}
\usepackage{amsmath,amssymb,amsfonts}
\usepackage{graphicx}
\usepackage{textcomp}
\usepackage{balance}
\usepackage{xcolor}
\usepackage{float}
\usepackage[utf8]{inputenc}
\usepackage{makecell}
\usepackage{multirow}
\usepackage{booktabs}
\usepackage{url}

\def\BibTeX{{\rm B\kern-.05em{\sc i\kern-.025em b}\kern-.08em
    T\kern-.1667em\lower.7ex\hbox{E}\kern-.125emX}}

\title{Trustworthiness Layer for Foundation Models in Power Systems: Application to N-k Contingency Screening}

\author{Antonio Alc\'antara, Spyros Chatzivasileiadis,~\IEEEmembership{Senior Member, IEEE}%
\thanks{A.~Alc\'antara is with the Dept.\ of Quantitative Methods, CUNEF Universidad, Madrid, Spain (e-mail: antonio.alcantara@cunef.edu). S.~Chatzivasileiadis is with the Dept.\ of Wind and Energy Systems, Technical University of Denmark, Kgs.\ Lyngby, Denmark (e-mail: spchatz@dtu.dk). This work was supported by the ERC Starting Grant VeriPhIED, Grant Agreement 949899 funded by the European Research Council.}%
}

\begin{document}
\begingroup
\allowdisplaybreaks

\maketitle

\begin{abstract}
We propose a model-agnostic trustworthiness layer that equips any foundation model (FM) for power systems with statistically valid prediction intervals. The layer offers two calibration approaches: (i)~stratified conformal prediction (SCP), which partitions residuals by contingency severity and grid element and comes with a formal coverage guarantee, and (ii)~kernel-weighted conformal prediction (KCP), which localizes the calibration to each test scenario, yielding tighter, scenario-adaptive bounds whose coverage we verify empirically. Using publicly released GridFM models as a guiding example, we demonstrate the framework on $N\!-\!k$ contingency screening for the IEEE 118-bus and a 500-bus system. Although the models are trained on $N\!-\!1$ data only, the true values stay within the bounds in about 90\% of cases at every contingency level up to $N\!-\!5$ on both systems. On IEEE-118, the layer raises detected overloads from about 81\% to 97\% at a precision comparable to that of DC power flow (DCPF), and cuts the share of insecure contingencies wrongly declared secure to 0.3\% or less, against 12\% for DCPF. On the 500-bus system, KCP cuts this share from 20--26\% (FM alone and DCPF) to 3.6--4.4\%. The layer also bounds bus voltages, which DCPF cannot screen. Adding under 0.1\,ms per scenario to the FM, it enables reliable large-scale security assessment beyond routine $N\!-\!1$ screening. An open-source implementation is publicly available.
\end{abstract}


\begin{IEEEkeywords}
Foundation Models, Conformal Prediction, Contingency Screening, Uncertainty Quantification, Power System Security.
\end{IEEEkeywords}

\section{Introduction}

Renewable integration and electrification drive unprecedented power system complexity, making security assessment a major challenge for system operators. Recent blackouts have shown that $N\!-\!1$ security is no longer sufficient; systems must be safeguarded against multiple simultaneous contingencies. While Transmission System Operators (TSOs) solve full AC Power Flow (ACPF) for routine $N\!-\!1$ screening, the combinatorial explosion of higher-order $N\!-\!k$ scenarios makes exhaustive ACPF computationally prohibitive~\cite{wood2013power}. Operators therefore resort to DC Power Flow (DCPF) for rapid large-scale screening~\cite{stott2009dc}. However, by neglecting reactive power and voltage variations, DCPF cannot detect voltage-limit violations~\cite{stott2009dc} or voltage instability~\cite{werkie2025power}, and its branch flows can be inaccurate~\cite{stott2009dc}.

Foundation Models (FMs) promise to bridge this gap, with the potential for large speed-ups over numerical solvers while preserving voltage and reactive-power information~\cite{hamann2024foundation}. In power grids, GridFM~\cite{hamann2024foundation} has been proposed as an FM built on Graph Neural Networks (GNNs) that learns the nonlinear AC physics from topological data, while task-specific GNN-based approaches have shown strong results in contingency analysis, generalizing from $N\!-\!1$ training scenarios to unseen topologies~\cite{nakiganda2023topology}. Where these speed-ups materialize, it becomes feasible to systematically evaluate thousands of higher-order $N\!-\!k$ contingencies with AC-level fidelity.

However, deploying ``black-box'' models in safety-critical infrastructure demands more than speed: operators need \emph{trustworthy} outputs with quantified uncertainty. An incorrect but plausible safe-state prediction during a genuine overload could trigger cascading failures. Prior ML-based contingency analysis work~\cite{nakiganda2023topology, yang2020power} has not addressed this gap with rigorous, distribution-free guarantees. Conformal Prediction (CP)~\cite{vovk2005algorithmic, shafer2008tutorial} provides such guarantees, but applying it to contingency screening is nontrivial due to the heterogeneous severity of $N\!-\!k$ events and element-wise nature of grid violations.

This work introduces a \textit{model-agnostic trustworthiness layer} for FMs in power systems\footnote{Open-source implementation, usable with GridFM models and with the predictions of any other model: \url{https://github.com/gridfm/gridfm-trustlayer}.}. The layer wraps any pre-trained FM with statistically valid prediction intervals via two complementary approaches: \emph{stratified} and \emph{kernel-weighted conformal prediction} (SCP, KCP). The goal of this paper is to use established statistical building blocks and tailor them to FM-based $N\!-\!k$ screening. Our contributions are: (i)~a post-hoc, retraining-free layer whose coverage guarantee and underlying assumptions are stated explicitly; (ii)~severity- and element-stratified calibration that remains valid when the FM extrapolates beyond the contingency orders seen during training; and (iii)~embedding-localized calibration which, unlike any global recalibration of the FM outputs, adapts the safety margin to each scenario and improves the precision--recall trade-off over SCP.

\section{Conformal Trustworthiness Layer}
\label{sec:method}

Consider a FM $f_\phi$ that, given a contingency scenario $x$ (comprising load, generation, and outages), predicts the AC bus voltages $\hat{y} = f_\phi(x)$. From $\hat{y}$, line loadings $\hat{L}_\ell$ are derived via the $\pi$-equivalent model, where $\hat{L}_\ell > 1$ indicates thermal overload. The trustworthiness layer computes an upper prediction bound $\hat{L}_\ell^{+}$ for each prediction such that the true line loading $L_\ell$ remains below this bound with a target coverage probability of $1-\alpha$, i.e., $\mathbb{P}(L_\ell \le \hat{L}_\ell^{+}) \ge 1-\alpha$. A line is flagged when $\hat{L}_\ell^{+} > 1.0$, ensuring conservative screening that minimizes missed violations (false negatives). The same framework applies to voltage magnitudes, with a lower bound for undervoltage (Section~\ref{sec:results}).

Both approaches start from a held-out calibration set $\mathcal{D}_{\mathrm{cal}}$ of $n$ scenarios $\{x_j, L_j\}_{j=1}^n$ and its residuals $R_{j,\ell} = L_{j,\ell} - \hat{L}_{j,\ell}$, where $L_{j,\ell}$ and $\hat{L}_{j,\ell}$ are the true and predicted loadings of line $\ell$ in scenario $j$. In plain terms, the layer records the FM errors on held-out scenarios with known AC solution, per line and contingency level; takes as safety margin the error exceeded in only a fraction $\alpha$ of them; and, at run time, adds this margin to the FM prediction, flagging the line if the result exceeds its rating. Margins can be recomputed for any $\alpha$ without retraining the FM.

\subsection{Approach~I: Stratified Conformal Prediction (SCP)}

Standard split CP~\cite{vovk2005algorithmic} uses a single global quantile, which is valid only on average over all scenarios: it under-covers severe contingencies and is over-conservative for routine ones. We address this by partitioning $\mathcal{D}_{\mathrm{cal}}$ into distinct subgroups (or `strata') based on contingency level $k$ (e.g. $N\!-\!1$, $N\!-\!2$) and line $\ell$:
{
\begin{equation}
\hat{q}^{(1-\alpha)}_{\ell,k} = R^{(k)}_{(\lceil (n_k+1)(1-\alpha)\rceil),\,\ell}
\label{eq:scp_quantile}
\end{equation}
}%
where $\mathcal{D}_{\mathrm{cal}}^{(k)}$ contains only the $n_k$ samples with $k$ simultaneous outages, and $R^{(k)}_{(r),\ell}$ is the $r$-th smallest residual of line $\ell$ among them; the $+1$ is the finite-sample correction of split CP~\cite{vovk2005algorithmic}. Since every scenario yields a residual for each line, each $(k,\ell)$ stratum contains all $n_k$ scenarios of level $k$. The upper bound becomes $\hat{L}_\ell^{+} = \hat{L}_\ell + \hat{q}^{(1-\alpha)}_{\ell,k}$.

{\textit{Assumptions and guarantee.}\; CP is distribution-free: it assumes nothing about the distribution of the scenarios or of the FM errors (the physics enters through the ACPF ground truth of the residuals), and only requires calibration and test scenarios to be exchangeable (e.g., i.i.d.\ draws of the same scenario generator). If this holds within each level $k$, then $\mathbb{P}(L_\ell \le \hat{L}_\ell + \hat{q}^{(1-\alpha)}_{\ell,k} \mid k) \ge 1-\alpha$ for every line $\ell$, marginally over calibration and test samples, not conditionally on a specific scenario. As quantiles are computed per stratum (Mondrian CP~\cite{vovk2005algorithmic}), exchangeability is needed neither across strata nor with the FM's training data: the guarantee holds for any test mix of contingency levels and for orders unseen during training, provided they are represented in $\mathcal{D}_{\mathrm{cal}}$, where poor extrapolation only widens the bounds. A different scenario distribution (e.g., a changed base topology) requires recalibration.\par}

\subsection{Approach~II: Kernel-Weighted Conformal Prediction (KCP)}

SCP applies the same margin to every scenario within a stratum. KCP weights the calibration residuals by their similarity to the current scenario, so that scenarios with larger FM errors receive wider margins, and well-predicted ones tighter margins.

\textit{Scenario representation and distance.}\; KCP requires a vector $z(x)$ characterizing each scenario $x$: a latent embedding of the FM or a feature vector of the input data. This is the only model-specific ingredient of the layer; SCP requires only the predictions. Each dimension of $z$ is standardized with its calibration mean and standard deviation. For the current test scenario $x_{\text{test}}$ (with representation $z_{\text{test}}$), squared Euclidean distances $d_j^2 = \|z_{\text{test}} - z_j\|_2^2$ are computed against all calibration scenarios $j$ in stratum $k$.

\textit{Adaptive bandwidth.}\; With weights $u_j = \exp(-d_j^2/h^2)$, the bandwidth $h^2$ is set per test point such that the effective sample size $n_{\mathrm{eff}} = (\sum_j u_j)^2/\sum_j u_j^2$ equals a target $\bar{n}_{\mathrm{eff}}$ (solved numerically), so that every bound rests on the same effective number of calibration scenarios.

\textit{Weighted quantile.}\; KCP replaces the SCP quantile~(\ref{eq:scp_quantile}) by a weighted one~\cite{tibshirani2019conformal}. Each calibration scenario $j$ of stratum $k$ gets the normalized weight $p_j = u_j/(1+\sum_i u_i)$; the remaining weight $p_0 = 1/(1+\sum_i u_i)$ belongs to the test scenario itself (distance zero, $u_0=1$), whose unknown residual is counted as $+\infty$. The margin of line $\ell$ is the smallest calibration residual at which the cumulative weight reaches $1-\alpha$:
\begin{equation}
\hat{q}_{\ell} = \min\Big\{ R_{j,\ell} \,:\, \textstyle\sum_{i:\,R_{i,\ell}\le R_{j,\ell}} p_i \;\ge\; 1-\alpha \Big\}.
\label{eq:kernel_weights}
\end{equation}
With equal weights ($u_j=1$), (\ref{eq:kernel_weights}) is exactly the SCP quantile~(\ref{eq:scp_quantile}); with kernel weights, residuals of similar scenarios count more. The test scenario's own weight keeps the bound conservative: the less calibration weight lies close to it, the larger $p_0$ and the margin. If $p_0>\alpha$ (the level is unreachable) or $n_{\mathrm{eff}} < n_{\mathrm{eff}}^{\min}$, KCP reverts to~(\ref{eq:scp_quantile}).

{\textit{Validity.}\; KCP is a localized conformal procedure~\cite{guan2023localized}: its kernel weights localize the calibration around the test scenario and are \emph{not} the density-ratio weights that make weighted CP valid under covariate shift~\cite{tibshirani2019conformal}. As implemented, KCP does not inherit the finite-sample guarantee of SCP; we regard SCP as the certified layer and KCP as a refinement whose coverage is verified empirically (Section~\ref{sec:results}).\par}

\section{Case Study: GridFM for N-k Screening}
\label{sec:results}

We demonstrate the trustworthiness layer on the publicly released GridFM models of~\cite{puech2026genco}, without retraining or fine-tuning: two power-flow checkpoints trained on $N\!-\!1$ contingency data, with 1.3M and 5.0M parameters (hidden dimensions 12 and 24, seed 0), denoted GFM-1.3M and GFM-5.0M. Calibration and test sets are generated with GridFM-datakit\footnote{Models: \url{https://huggingface.co/gridfm}; data generation: \url{https://github.com/gridfm/gridfm-datakit}.} for $N\!-\!1$ to $N\!-\!5$ contingencies with the same settings and independent draws (19{,}913 calibration and 199{,}076 test scenarios); $N\!-\!2$ to $N\!-\!5$ thus lie outside the training distribution, yet are covered by the SCP guarantee. For KCP, $z(x)$ is the mean over buses of the final-layer bus embeddings of the FM (96 and 192 dimensions), with $\bar{n}_{\mathrm{eff}} = 600$ and $n_{\mathrm{eff}}^{\min} = 5$; $\bar{n}_{\mathrm{eff}}$ was selected on a separate tuning set of 20k scenarios. As a second, larger system, we use the 500-bus PGLib case \texttt{case500\_goc} with its released checkpoints of the same two sizes, 19{,}978 calibration and 49{,}939 test scenarios, and the same layer settings.

Our analysis focuses on the trustworthiness layer, not the underlying FM: GridFM serves as a guiding example, and the layer applies to any model predicting continuous system states, DCPF included (DCPF+SCP). We compare original and calibrated predictions (Point vs.\ +SCP/+KCP), with DCPF, the tool operators use for fast large-scale screening~\cite{stott2009dc}, as the classical baseline in two variants: the standard DCPF (DC branch flows against ratings) and DCPF$^{\ast}$ (AC branch equations at the DC state, with flat voltage magnitudes at PQ buses).

\textit{Metrics.}\; Recall, the critical safety metric, is the fraction of true overloads that are flagged, and precision the fraction of flags that are true overloads (DCPF and the point estimates flag $\hat{L}_\ell>1$). A contingency without flagged line is declared secure, and thus missed if it is insecure (has an overload); one with a flagged line is passed to ACPF, which reveals all its violations. A lower bound $\hat{L}^{-}_\ell$, from~(\ref{eq:scp_quantile}) with rank $\lfloor (n_k+1)\alpha\rfloor$, splits flagged contingencies into \emph{confirmed} insecure ($\hat{L}^{-}_\ell>1$ for some line) and \emph{uncertain}; only the latter need ACPF to be classified. Coverage is not recall: 90\% coverage means that 90\% of all line--scenario pairs, most of them far from an overload, lie below their bound. As a covered overload is always flagged, recall is lower-bounded by $\mathbb{P}(L_\ell\le\hat{L}^{+}_\ell \mid L_\ell>1)$, which is not guaranteed (it would require label-conditional CP~\cite{vovk2005algorithmic}); recall is thus an empirical metric.

\begin{table}[t]
\caption{Recall (Precision) for Line Congestion Screening by $N\!-\!k$ Level on IEEE-118, and Overall on the 500-Bus System ($1-\alpha = 0.9$). DCPF$^{\ast}$: AC Equations at the DC State}
\label{tab:nk_results}
\centering
{\renewcommand{\arraystretch}{0.92}
\resizebox{\columnwidth}{!}{%
\begin{tabular}{lcccccccccc}
\toprule
& & & & \multicolumn{3}{c}{\textbf{GFM-1.3M}} & & \multicolumn{3}{c}{\textbf{GFM-5.0M}} \\
\cmidrule(lr){5-7} \cmidrule(lr){9-11}
& \textit{\textbf{DCPF}} & \textit{\textbf{DCPF$^{\ast}$}} & & \textbf{Point} & \textbf{+SCP} & \textbf{+KCP} & & \textbf{Point} & \textbf{+SCP} & \textbf{+KCP} \\
\midrule
$N\!-\!1$ & \makecell{0.416\\{\scriptsize(0.607)}} & \makecell{0.624\\{\scriptsize(0.120)}} & & \makecell{0.703\\{\scriptsize(0.735)}} & \makecell{0.959\\{\scriptsize(0.637)}} & \makecell{0.953\\{\scriptsize(0.646)}} & & \makecell{0.698\\{\scriptsize(0.749)}} & \makecell{0.956\\{\scriptsize(0.653)}} & \makecell{0.954\\{\scriptsize(0.660)}} \\
$N\!-\!2$ & \makecell{0.474\\{\scriptsize(0.705)}} & \makecell{0.678\\{\scriptsize(0.150)}} & & \makecell{0.783\\{\scriptsize(0.830)}} & \makecell{0.972\\{\scriptsize(0.705)}} & \makecell{0.970\\{\scriptsize(0.715)}} & & \makecell{0.790\\{\scriptsize(0.846)}} & \makecell{0.973\\{\scriptsize(0.729)}} & \makecell{0.973\\{\scriptsize(0.739)}} \\
$N\!-\!3$ & \makecell{0.511\\{\scriptsize(0.768)}} & \makecell{0.713\\{\scriptsize(0.173)}} & & \makecell{0.823\\{\scriptsize(0.876)}} & \makecell{0.973\\{\scriptsize(0.747)}} & \makecell{0.973\\{\scriptsize(0.759)}} & & \makecell{0.842\\{\scriptsize(0.893)}} & \makecell{0.979\\{\scriptsize(0.773)}} & \makecell{0.979\\{\scriptsize(0.783)}} \\
$N\!-\!4$ & \makecell{0.546\\{\scriptsize(0.814)}} & \makecell{0.741\\{\scriptsize(0.197)}} & & \makecell{0.845\\{\scriptsize(0.903)}} & \makecell{0.970\\{\scriptsize(0.769)}} & \makecell{0.971\\{\scriptsize(0.781)}} & & \makecell{0.867\\{\scriptsize(0.917)}} & \makecell{0.975\\{\scriptsize(0.800)}} & \makecell{0.978\\{\scriptsize(0.810)}} \\
$N\!-\!5$ & \makecell{0.574\\{\scriptsize(0.847)}} & \makecell{0.765\\{\scriptsize(0.217)}} & & \makecell{0.855\\{\scriptsize(0.921)}} & \makecell{0.968\\{\scriptsize(0.782)}} & \makecell{0.972\\{\scriptsize(0.789)}} & & \makecell{0.883\\{\scriptsize(0.934)}} & \makecell{0.975\\{\scriptsize(0.818)}} & \makecell{0.979\\{\scriptsize(0.828)}} \\
\midrule
\textbf{All} & \makecell{\textbf{0.509}\\{\scriptsize(0.756)}} & \makecell{\textbf{0.709}\\{\scriptsize(0.170)}} & & \makecell{\textbf{0.807}\\{\scriptsize(0.859)}} & \makecell{\textbf{0.968}\\{\scriptsize(0.729)}} & \makecell{\textbf{0.968}\\{\scriptsize(0.740)}} & & \makecell{\textbf{0.822}\\{\scriptsize(0.874)}} & \makecell{\textbf{0.972}\\{\scriptsize(0.756)}} & \makecell{\textbf{0.973}\\{\scriptsize(0.766)}} \\
\midrule
\makecell[l]{\textbf{All,}\\\textbf{500-bus}} & \makecell{\textbf{0.594}\\{\scriptsize(0.579)}} & \makecell{\textbf{0.771}\\{\scriptsize(0.005)}} & & \makecell{\textbf{0.688}\\{\scriptsize(0.591)}} & \makecell{\textbf{0.871}\\{\scriptsize(0.438)}} & \makecell{\textbf{0.909}\\{\scriptsize(0.414)}} & & \makecell{\textbf{0.721}\\{\scriptsize(0.636)}} & \makecell{\textbf{0.891}\\{\scriptsize(0.498)}} & \makecell{\textbf{0.926}\\{\scriptsize(0.475)}} \\
\bottomrule
\end{tabular}%
}}
\end{table}

\begin{table}[t]
\caption{Coverage of the Bound and Screening Outcome at $1-\alpha = 0.9$ (GFM-5.0M). Missed: \% of Insecure Contingencies; Flagged: \% of All}
\label{tab:diag}
\centering
{\footnotesize\setlength{\tabcolsep}{3.5pt}\renewcommand{\arraystretch}{0.9}
\begin{tabular}{@{}clccccc@{}}
\toprule
& & \multicolumn{3}{c}{\textbf{Coverage of the bound (\%)}} & \multicolumn{2}{c}{\textbf{Contingencies (\%)}} \\
\cmidrule(lr){3-5} \cmidrule(lr){6-7}
& & $N\!-\!1$ & $N\!-\!5$ & Worst stratum & Missed & Flagged \\
\midrule
\multirow{4}{*}{\rotatebox{90}{\scriptsize IEEE-118}} & Point & 49.6 & 46.3 & 36.7 & 5.5 & 90.7 \\
& \quad +Global CP & 95.3 & 83.7 & 81.2 & 0.22 & 97.9 \\
& \quad +SCP & 90.1 & 89.7 & 88.2 & 0.28 & 97.9 \\
& \quad +KCP & 90.6 & 89.9 & 88.0 & 0.30 & 97.7 \\
\midrule
\multirow{4}{*}{\rotatebox{90}{\scriptsize 500-bus}} & Point & 52.5 & 49.5 & 25.6 & 20.0 & 22.3 \\
& \quad +Global CP & 91.6 & 87.6 & 81.9 & 5.0 & 32.0 \\
& \quad +SCP & 90.0 & 89.9 & 87.9 & 5.0 & 32.2 \\
& \quad +KCP & 91.6 & 90.4 & 87.2 & 3.6 & 33.5 \\
\bottomrule
\end{tabular}}
\end{table}

\textit{Detection.}\; Table~\ref{tab:nk_results} reports recall and precision by $N\!-\!k$ level. The point estimates already outperform DCPF in recall and precision, but miss about one in five overloads. SCP and KCP raise recall to 0.968--0.973 overall and to at least 0.953 at every level, including $N\!-\!5$, the farthest from the training data (0.968--0.979), at a precision of 0.729--0.766, comparable to that of DCPF. DCPF$^{\ast}$, which inflates the loading, reaches a recall of 0.709 at a precision of only 0.170.

\textit{Validation.}\; Table~\ref{tab:diag} checks coverage and screening on both systems (GFM-1.3M is similar). SCP and KCP cover 89.7--90.6\% at every level, including $N\!-\!5$. Over the 930 $(k,\ell)$ strata of IEEE-118, each with 3{,}544--4{,}268 calibration samples, SCP ranges from 88.2\% to 91.6\% for both models, as expected from finite-sample variability, and KCP falls back to SCP in at most 0.14\% of the scenarios. Stratification is essential: one quantile per line pooled over all levels (Global CP) also covers 90\% overall, with similar recall and precision, but 95--96\% at $N\!-\!1$ and only 83--84\% at $N\!-\!5$ on IEEE-118, as the FM errors grow with $k$, whereas SCP holds the target at every level. As 92.7\% of the IEEE-118 contingencies are insecure, nearly all are flagged, but SCP confirms 67--68\% as insecure (0.9\% wrongly), leaving 30--32\% uncertain. The layer misses 0.23--0.30\% of the insecure ones (point estimates: 5.5--5.8\%; DCPF: 12.1\%) and detects 98.8--99.4\% of the severe overloads ($L_\ell>1.1$).

\textit{Comparison with threshold tuning.}\; Fig.~\ref{fig:pr_tradeoff} shows the precision--recall trade-off for different targets $1-\alpha$ and for classical threshold tuning (gray), which flags a line if $\hat{L}_\ell > \tau$ for a uniform $\tau$. Any global monotone recalibration $g$ of $\hat{L}_\ell$ (e.g., isotonic regression) flags a line iff $\hat{L}_\ell$ exceeds a fixed threshold, i.e., is a point on this curve, whereas the margins of SCP and KCP depend on the stratum and the scenario. Both lie above the threshold-tuning curves, although $\tau$ is swept on the test set, i.e., knowing the true overloads: at the same recall of about 0.97, their precision is 3.5--5 points higher, and at the same share of flagged contingencies they miss about half as many insecure ones (0.23--0.30\% vs.\ 0.44--0.58\%). KCP has the highest precision at every coverage level, about one point above SCP at equal recall.

\textit{Calibrated DCPF.}\; The layer applies to any predictor, DCPF included (purple in Fig.~\ref{fig:pr_tradeoff}): it restores coverage to 90\% and raises recall to 0.965, but with margins 10--13 times wider than with the FMs, hence a precision of only 0.55. The guarantee is the same; only its cost depends on accuracy.

\textit{500-bus system.}\; Only 20.6\% of the contingencies are insecure, so screening avoids most ACPF solves. SCP and KCP again cover 89.8--91.6\% at every level, but as the FMs underestimate rare high loadings, SCP reaches a recall of only 0.871--0.891 (Table~\ref{tab:nk_results}). KCP, which widens the margin where the FM is less accurate, restores 0.909--0.926, with a precision 0.03--0.05 above SCP at equal recall and within about 0.02 of the best test-set threshold. It flags 33--36\% of the contingencies but misses only 3.6--4.4\% of the insecure ones (point estimates and DCPF: 20--26\%). DCPF+SCP comes close (recall 0.877, precision 0.519) but misses 6.0\% and detects only 86\% of the severe overloads (FMs: 93.5--97.8\%).

\textit{Voltage limits.}\; Applied unchanged to IEEE-118 bus voltages (PQ buses; limits 0.94/1.06\,pu), lower and upper bounds cover 89.7--90.7\% at every level. Of the 6.8\% of contingencies with undervoltage, the point estimates miss 14--16\%, SCP and KCP 3.2--5.7\%, flagging $\approx$8.5\% of all contingencies, with a precision 0.05--0.08 above the best test-set threshold at equal recall. Overvoltage recall rises from 0.74--0.78 to 0.97, at the threshold's precision. DCPF (flat voltages) detects neither.

\begin{figure}[t]
    \centering
    \includegraphics[width=\linewidth]{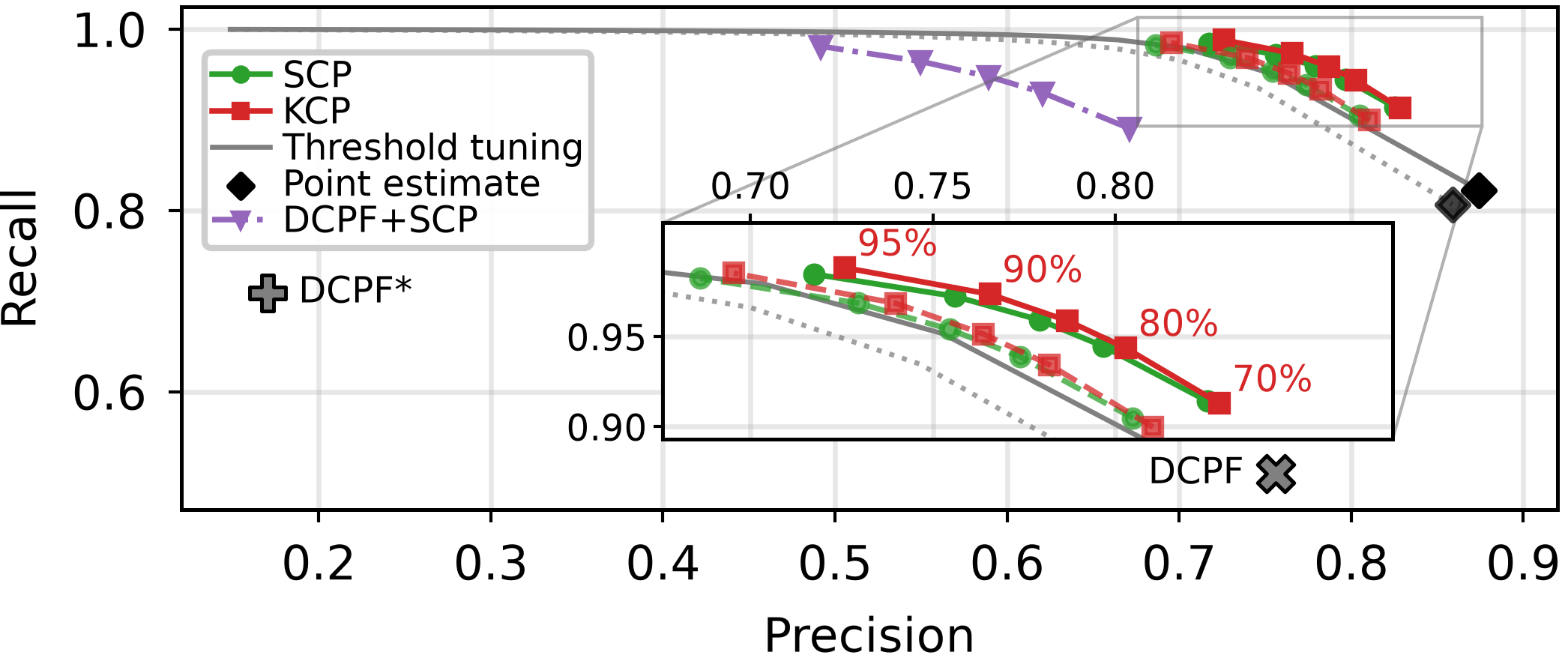}
    \caption{Precision--Recall trade-off on IEEE-118 at different coverage $(1-\alpha)$ levels; solid: GFM-5.0M, dashed/dotted: GFM-1.3M; labels: KCP targets.}
    \label{fig:pr_tradeoff}
\end{figure}

\textit{Computational cost.}\; The layer is light: on an Apple M4 Pro, SCP adds under 2\,$\mu$s per scenario and KCP 24--28\,$\mu$s (500-bus: 69--82\,$\mu$s). Calibration is a small, one-off offline step: with 1,000 scenarios, SCP keeps 90\% average coverage on both systems and loses at most 0.01 of precision; generating 20k scenarios took 5.1\,min (500-bus: 24\,min) on 10 CPUs. An IEEE-118 scenario takes 1.30\,ms with ACPF and 0.57\,ms with DCPF (PowerModels, one process), and 1.1/2.1\,ms (GPU) or 2.2/4.6\,ms (CPU) with GFM-1.3M/5.0M in batches of 256; the FM speed advantage grows with grid size, to about 28--29$\times$ over ACPF on 2,000- and 10,000-bus systems~\cite{puech2026genco}.

\balance
\section{Conclusion}
We developed a post-hoc, model-agnostic trustworthiness layer that equips any power-system FM with statistically valid bounds, without retraining. The SCP variant gives finite-sample coverage per contingency level and grid element, and KCP adapts the margin to each scenario. With two released GridFM models trained on $N\!-\!1$ data only, the layer kept coverage at about 90\% at every level up to $N\!-\!5$ on two systems, for line loadings and bus voltages. It cut missed insecure contingencies to 0.3\% or less on IEEE-118 (12\% with DCPF) and to 3.6--4.4\% on the 500-bus system (20--26\% for the point estimates and DCPF), adding under 0.1\,ms per scenario. As the guarantee does not depend on the predictor, the layer also calibrates DCPF; a more accurate model only tightens the bounds. Future work will address distribution shifts, guarantees on recall and at system level, angle limits, and larger grids, where FMs are fastest.

\bibliographystyle{IEEEtran}

\endgroup
\end{document}